\documentclass[12pt]{article}
\usepackage{epsfig}

\def\beq{\begin{equation}}
\def\eeq#1{\label{#1}\end{equation}}
\def\eeqn{\end{equation}}

\def\beqa{\begin{eqnarray}}
\def\eeqa#1{\label{#1}\end{eqnarray}}
\def\eeqan{\end{eqnarray}}

\let\bar=\overbar

\def\Dslash{\not{\hbox{\kern-4pt $D$}}}
\def\dslash{\not{\hbox{\kern-2pt $\del$}}}

\def\msb{{\bar{\ssstyle M \kern -1pt S}}}

\usepackage{fancyhdr,graphicx}
\def\Title#1{\begin{center} {\Large {\bf #1} } \end{center}}

\begin{document}

\Title{Chiral Transition Within Effective Quark Models under Strong Magnetic Fields}

\bigskip\bigskip


\begin{raggedright}

{\it Andr\'e Felipe Garcia and Marcus Benghi Pinto\index{Vader, D.}\\
Departamento de F\'isica\\
Universidade Federal de Santa Catarina\\
88040-900 Campus Universit\'ario\\
Florian\'opolis, SC\\
Brazil\\
{\tt Email: andregarcia000@gmail.com}}
\bigskip\bigskip
\end{raggedright}

\section{Introduction}

A widely studied subject within the context of strong interactions in high energy physics is that of the QCD phase diagram. After it had became clear that hadrons are composed by confined quarks and gluons it was suggested that they might undergo a phase transition at high temperature or density, becoming a deconfined plasma, the so called ``quark-gluon plasma". This transition has significant experimental implications (some of them being tested in modern accelerators such as the LHC, RHIC, etc), not to mention the description of the early stages of the universe and the matter inside neutron stars.

In the recently years it has been argued that spectators in heavy ion collisions are responsible for creating a strong magnetic field that could play an important role in the QCD phase transition.
In this work we use the SU(2) Nambu--Jona-Lasinio (NJL) model in order to study the chiral transition in quark matter subject to a strong magnetic field. We show some of our results presented in Ref. \cite{nos}, involving the breaking of chiral symmetry and its restoration at finite temperature and density.

\section{Evaluation of the Landau Free Energy}

We aim to use de Nambu-Jona-Lasinio model, Ref. \cite{Nambu}, in the SU(2) version in order to study the chiral transition in quark matter with three color degrees of freedom. In the presence of a magnetic field, the lagrangian of the model is

\begin{equation} \label{lagrangianaB}
{\cal L} =
\bar \psi_f \left( i \not\!\partial -q_f \gamma_{\mu}A^{\mu}-m \right) \psi_f +
G\left[(\bar \psi_f \psi_f)^2 + (\bar \psi_f i \gamma_5 \vec{\tau} \psi_f)^2 \right] -\frac{1}{4}F^{\mu\nu}F_{\mu\nu}\, ,
\end{equation}

\noindent
where $\psi_f$ represents the quark field with flavor $f$, $q_f$ is the quark electric charge, $A^{\mu}$ is the vector potential, $m$ is the quark bare mass (assumed to be equal for both \textit{up} and \textit{down} quarks), $G$ is the coupling constant, $\vec\tau$ the Pauli matrices and $F^{\mu\nu}=\partial^{\mu}A^{\nu}-\partial^{\nu}A^{\mu}$.

Following Ref. \cite{Debora} we can write the Landau free energy in mean field approximation as:

\begin{equation} \label{Ftot}
\mathcal{F}=\frac{(M-m)^2}{4G} + \mathcal{F}_{vac}+\mathcal{F}_{mag}+\mathcal{F}_{med}
\end{equation}

\noindent where

\begin{equation} \label{Fvac}
\mathcal{F}_{vac}=-2N_cN_f \int \frac{d^3p}{(2\pi)^3}E_p
\end{equation}

\begin{equation} \label{Fmag}
\mathcal{F}_{mag}=-\sum_{f=u}^d \frac{N_c(|q_f|B)^2}{2\pi^2}\left\{\zeta'[-1,\,x_f]-\frac{1}{2}[x_f^2-x_f]\textrm{ln}(x_f)+\frac{x_f^2}{4} \right\}
\end{equation}

\begin{eqnarray} \label{Fmed}
\mathcal{F}_{med}=-\frac{N_c}{2\pi}\sum_{f=u}^d\sum_{k=0}^{\infty}\alpha_k(|q_f|B)\int_{-\infty}^{\infty} \frac{dp_z}{2\pi}\left\{T\textrm{ln}[1+\textrm{e}^{-[E_{p,\,k}(B)+\mu]/T}]+ \right . \nonumber \\
\left . T\textrm{ln}[1+\textrm{e}^{-[E_{p,\,k}(B)-\mu]/T}]\right\}
\end{eqnarray}

\noindent are the vacuum, magnetic and medium contributions, respectively.
In Eq. (\ref{Fvac}) $N_c=3$ and $N_f=2$ are the color and flavor degrees of freedom, respectively,  and the integral is carried out up to a cutoff $\Lambda=590\rm{MeV}$. In Eq. (\ref{Fmag}) $x_f=M^2/(2|q_f|B)$ and $\zeta'[-1,\,x_f]=d\zeta(z,\,x_f)/dz|_{z=-1}$ with $\zeta(z,\,x_f)$ being the Riemann-Hurwitz function. We also fix the coupling constant as $G=2.44/\Lambda^2$. Finally, in Eq. (\ref{Fmed}) we have $E_{p,\,k}(B)=\sqrt{p_z^2+2k|q_f|B +M^2}$, with $k$ being the Landau Levels.




\section{Results}

\begin{figure}[htb]
\begin{center}
\epsfig{file=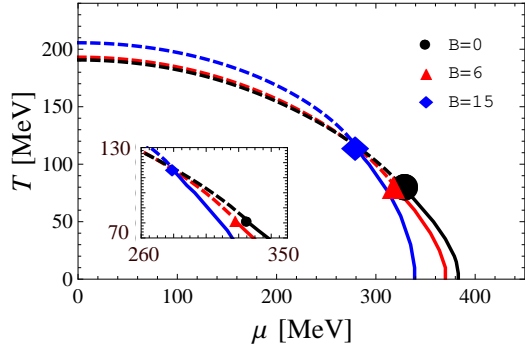,height=1.8in}
\caption{Phase diagram for the NJL model in the $T-\mu$ plane for different values of the magnetic field in units of $m_{\pi}^2/e$. The continuous/dashed lines represent first order phase transitions/crossovers. The critical points are represented by the solid symbols.}
\label{fig:phasediagram}
\end{center}
\end{figure}

The restoration of chiral symmetry takes place at finite temperature and/or chemical potential. The phase diagram is shown in Fig. \ref{fig:phasediagram}, where we have plotted the first order phase transition lines and the pseudo-temperature crossover lines for different values of the magnetic field. We observe that the critical point is shifted depending on the value of $B$ and that the crossover pseudo-temperature at $\mu=0$ always increases with $B$. Recently, lattice QCD results suggest that this may not be the case, leaving open the question of whether the magnetic field increases or not the crossover pseudo-temperature.

All the relevant thermodynamical quantities can be readily obtained by recalling that the free energy, evaluated at the mass value which satisfies the gap equation, gives the negative of the pressure, $\mathcal{F}(M)=-P$. Then, the net quark number density is obtained from $\rho=dP/d\mu$, and the entropy density from $s=dP/dT$ while the energy density is $\varepsilon=-P+Ts+\mu\rho$.

\begin{figure}[htb]
\begin{center}
\epsfig{file=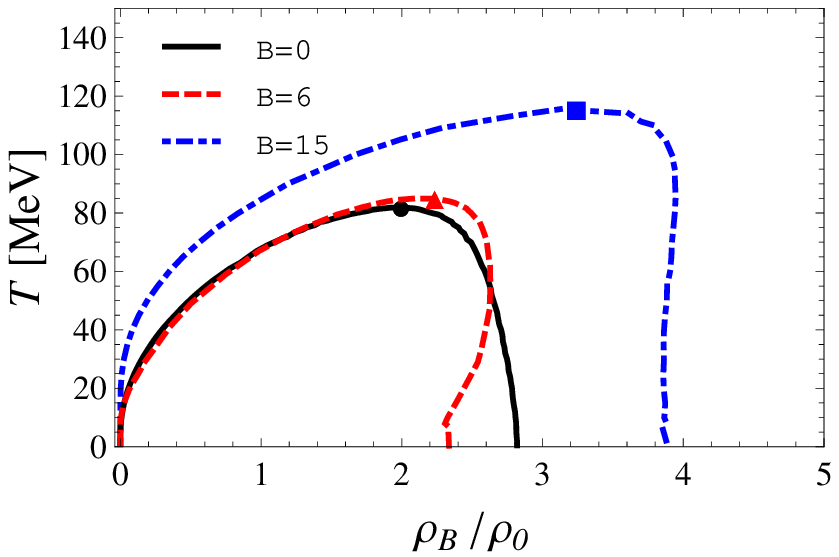,height=1.7in,angle=0}
\epsfig{file=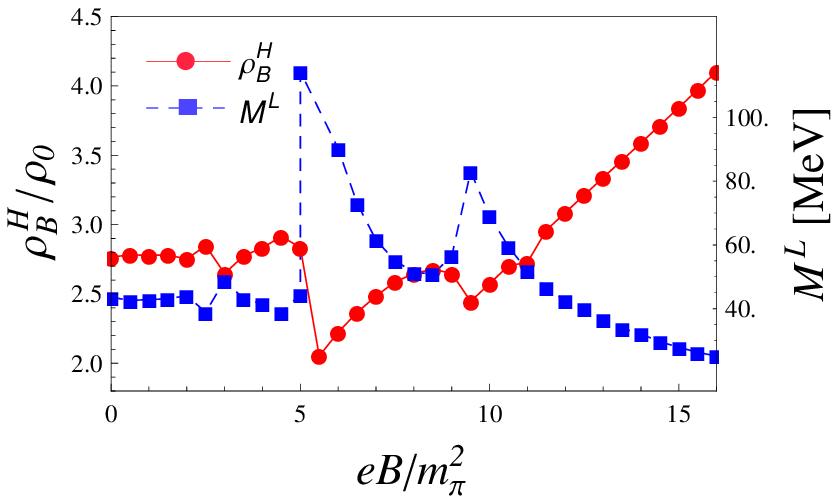,height=1.8in,angle=0}
\caption{(Left) The phase coexistence diagram in the $T-\rho_B$ plane for different values of the magnetic field in units of $m_{\pi}^2/e$. (Right) The oscillatory behavior of $\rho_B$ and $M$ at the restored phase ($T=0$).}
\label{fig:Trho}
\end{center}
\end{figure}

An interesting feature arises when we look at the coexistence diagram in the $T-\rho_B$ plane, Fig. \ref{fig:Trho} (left panel). For low temperature we see that the higher value of $\rho_B$ for $B\neq0$ oscillates around the $B=0$ line. This is due to the filling of the Landau levels, since that in the limit $T\rightarrow 0$, $\rho_B$ behaves like

\begin{equation}
 \rho_B (\mu,B)= \theta (k_F^2)\sum_{f=u}^d \sum_{k=0}^{k_{max}} \alpha_k \frac{|q_f| B N_c}{6\pi^2}k_F \,\,,
\end{equation}
where $k_F=\sqrt{\mu^2-2|q_f|kB-M^2}$, $\theta$ is the Heaviside function and
\begin{equation}
k_{max} = \frac{\mu^2 - M^2}{2|q_f|B} \, ,
\end{equation}

\noindent or the nearest integer.

The right panel of Fig. \ref{fig:Trho} shows the oscillatory behavior of $\rho_B$ (left axis) as well as the opposite oscillatory behavior of the constituent quark mass at the symmetry restored phase (right axis) as a function of the magnetic field. The origin of the oscillations in these quantities can be traced back to the fact that $k_{max}$ (the upper Landau level filled) decreases as the magnetic field increases.

\section{Conclusion}

We studied the effects of a strong magnetic field in the chiral transition using the two flavor Nambu-Jona-Lasinio model. We found that the critical end point of the first order transition is shifted towards higher $T$ and lower $\mu$ as $B$ increases. We also found that the crossover pseudo-temperature at $\mu=0$ always increases with $B$. This is in contrast with recent QCD lattice calculations, that suggest that the crossover pseudo-temperature may actually decrease with $B$. At the coexistence diagram in the $T-\rho_B$ plane we found that for low $T$ the coexistence lines in the higher density branch for $B\neq 0$ oscillates around the $B=0$ curve. We explained this feature in terms of the filling in the Landau levels at different $B$.

\bigskip
AFG would like to thank the organizing committee and the PGFSC-UFSC for financial support.

\end{document}